\documentclass{article}[15pt]

\usepackage{amsmath,epsfig,color,calc,rotating}
\usepackage{wrapfig}

\begin{document}

\title{Copy-Number-Variation and Copy-Number-Alteration Region 
Detection by Cumulative Plots
\vspace{0.2in}
\author{
Wentian Li\footnote{Corresponding author: wli@nslij-genetics.org},
Annette Lee, Peter K Gregersen  \\
{\small \sl The Robert S. Boas Center for Genomics and Human Genetics, 
The Feinstein Institute for Medical Research}\\
{\small \sl for Medical Research, North Shore LIJ Health System,
 Manhasset, 350 Community Drive, NY 11030, USA.}\\
}
\date{preprint version of {\em BMC Bioinformatics}, 10(suppl 1):S67  (2009) }
}
\maketitle

\begin{abstract}
        {\bf Background:} 
Regions with copy number variations (in germline cells) or copy number alteration 
(in somatic cells) are of great interest for human disease gene mapping and cancer studies.
They represent a new type of mutation and are larger-scaled than the single nucleotide 
polymorphisms. Using genotyping microarray for copy number variation detection has become 
standard, and there is a need for improving analysis methods.
        {\bf Results:} We apply the cumulative plot to the detection
of regions with copy number variation/alteration, on samples taken from a chronic 
lymphocytic leukemia patient. Two sets of whole-genome genotyping of 317k single 
nucleotide polymorphisms, one from the normal cell and another from the cancer cell, 
are analyzed.  We demonstrate the utility of cumulative plot in detecting a 9Mb 
(9 $\times 10^6$ bases) hemizygous deletion and 1Mb homozygous deletion on chromosome 13. 
We also show the possibility to detect smaller copy number variation/alteration regions below 
the 100kb range.
        {\bf Conclusions:} As a graphic tool, the cumulative plot
is an intuitive and a scale-free (window-less) way for detecting copy number
variation/alteration regions, especially when such regions are small.

\end{abstract}

\newpage


\large

\section*{Background}

\indent

Most efforts in genetic mapping of human diseases focus on
single-nucleotide-polymorphism (SNP): individual nucleotide base
that may differ from one person to another. If the cause of
a polymorphism is due to diverging paths in population genetic history,
such as in multiple ethnic groups, it can be used as an ancestry 
or ethnic identity marker\cite{rosenberg}. If the polymorphism is 
a functional mutation (non-synonymous or promoter-region polymorphism)\cite{botstein} 
underlying a human disease, then it is the focus of
attention in case-control genetic analyses\cite{wli-bi}.

A new type of genetic polymorphism emerged recently as another
source of mutation that may lead to human diseases: the copy number
variation (CNV) (for literature on CNV, see an online bibliography \cite{cnvbib}). 
Local duplication and deletion events occuring at kb ($10^3$ bases) or 
Mb ($10^6$ bases) scales are the cause of CNV.
If these events occurred in prior generations, CNV can be treated
as a genetic marker whose transmission might be traced in studying 
the disease-status correlation. These events can also occur in the current
generation, as {\sl de novo} mutations.

Similar duplication and deletion events also occur in somatic
cells, leading to copy number alteration (CNA). Besides
the link between CNA and cancers studied before\cite{cgh}, an early 
CNV-disease association was reported on Charcot-Marie-Tooth 
disease\cite{lupski95}, in inherited neurological disorder.
In the past year or two, the number of reports on association of CNV
with human diseases increased dramatically, especially for 
psychiatric disorders such as Schizophrenia\cite{sutrala,walsh,xu},
bipolar\cite{wilson}, and for brain developmental disorder 
such as Autism\cite{sebat,agp,weiss,marshall,eichler}. These
diseases have long been evading genetic dissection, and the CNV
link offers new optimism for our ultimate understanding of
these diseases.

The technology for CNV detection evolves from Mb-level comparative genomic
hybridization (CGH) to higher-resolution array-based CGH\cite{carter}.
Genotyping array whose original goal is to genotype individual SNPs, has 
increasingly been used for CNV detection\cite{huang,nannya,newman,peiffer}. 
There are two relevant pieces of information from a genotyping array data
for the purpose of CNV detection. 

The first is the ratio of intensity reading of alleles for a sample 
to that from a reference group of normal samples. If the ratio is larger 
than 1, there are more copies of piece of DNA in the sample 
than normal (which is 2 copies). If the ratio is less than 1, it indicates 
a deletion.  The second signal is the genotype. Deletion of one
of the chromosomes leads to a run of homozygosity for all SNPs in the region,
though run of homozygosity can also be due to inbreeding\cite{morton,simon}.
The homozygosity property of one-copy deletion is well exploited in detecting 
loss-of-heterozygosity in CNA of cancer cells\cite{lin}.

CNV detection using genotype microarray data relies on these two sequences:
if the intensity ratio deviates from the normal value of 1 for a chromosome
region with a consistent value, it can be a CNV region. Similarly,
if a run of homozygosity is observed in a region, it could indirectly
indicate a copy-number deletion. A CNV region detection is
more convincing if CNV signals exhibited by both sequences overlap in a
common region.

Methods for calling CNV regions can be roughly classified into two types.
The first type is straightforward: a CNV detection is claimed when the 
log-ratio value is significantly deviated from 0\cite{vermeesch}. The problem 
with this method is that the threshold for calling CNV varies greatly from platform to
platform, from study to study, and a comparative investigation is urgently
needed\cite{carter}. The second type uses hidden Markov models (HMM),
where the underlying CNV status is the hidden variable, and the 
log-ratio and genotype sequences are the two observed 
variables\cite{zhao,fridlyand,cli,colella,wang,cahan}.
One advantage of the HMM framework is that it can incorporate information
from both sequences at once.

When the parameter settings in a HMM are fixed, HMM is a stationary 
(homogeneous) process along a chromosome. There is one parameter in HMM
which controls the
transition probability from the (hidden) CNV state to non-CNV state.
That parameter can also be transformed to the characteristic size
for CNV region\cite{colella}. What if the CNV regions do not have
a characteristic size, or equivalently, the length distribution is not exponential?
In that case, CNV-calling methods that do not require stationarity
are preferred.

The guanine-cytosine content (GC\%) in DNA sequences has been a
focus of non-stationary, non-Markov, long-range-correlated modeling 
for more than twenty years\cite{fickett,likaneko,melo}.
It is well acknowledged that the hierarchical pattern of
GC\%-domains within GC\%-domains is possible\cite{li01,li02}. 
In order to detect both small and large GC-homogeneous 
domains, one applies methods that do not preset a characteristic scale.
One such method is the recursive segmentation
that adopts a divide-and-conquer approach\cite{pedro}. Another is
the cumulative plot.

Cumulative plot is a graphic display of sequence information such
that trend in a region becomes more visible and obvious. It is
a window-less method because no characteristic scale needs to
be specified, although a window can be imposed to a plot when
all patterns within certain length scale are to be ignored. In
DNA sequence context, such cumulative plots were called 
``DNA walk"\cite{berthelsen,peng} or ``Z curve"\cite{z94,z03-1}.
The cumulative plot has also been widely used for detection of
replication origin\cite{grigo,freeman}.  To our knowledge, 
cumulative plots have not been applied to CNV/CNA detection. 
The purpose of this paper is not to provide a comprehensive comparison of 
various CNV/CNA-calling methods, but limited to the presentation
and illustration of this new approach.

\section*{Results and discussion}

\indent

Since our method applies equally to CNV and CNA data, here
we examine the CNA pattern in a cancer patient with chronic lymphocytic
leukemia (CLL)\cite{cll2}. DNA samples from the patient's normal cell and
that from the cancer cell are obtained and genotyped with 317,000 SNPs
genomewide. Figure 1 shows the log-ratio and $\theta$ sequences
(see Methods) for chromosome 13, where a 9Mb CNA region (deletion) 
in the cancer cell is clearly visible. A deletion region is characterized 
by a drop in log-ratio value, and an absence of heterozygosity. 
Our goal is to capture the same information using cumulative plots.

The left panel of Figure 2 shows the two cumulative plots corresponding
to log-ratio sequence and homozygosity indicator sequence $h$, respectively. In
the simplest version, at each new SNP, the curve moves up or down
by an amount equal to the log-ratio value of that SNP, or by the
presence of a homozygote (+1) and a heterozygote ($-1$). 

For a deletion region, the log-ratio value is consistently negative, 
and the first cumulative plot shows a drop; and genotype is consistently homozygous
(also called run of homozygosity (ROH)), and the second cumulative plot 
shows a jump. However, from Figure 2 (left), even outside the CNA region, the
first (second) cumulative plot continues to go down (up), reflecting 
a global abundancy of negative log-ratio over positive one
(homozygotes over heterozygotes). 

To remove the global or chromosome-wide average, we redraw a 
detrended cumulative plot (right panel of Figure 2) where the 
linear trend from the normal cell is subtracted from the two 
cumulative plots. If the difference of global trends between
the cancer and normal cell is an artifact, e.g., the poor DNA quality
in cancer cell that leads to higher missing rate for genotype calling,
thus seemingly lower heterozygote frequency, then the
normal and cancer cumulative plots should be detrended separately.
Without such an evidence, we use the linear trend in normal
cell to detrend both samples to highlight the difference
between the two.

Cumulative plots can be customized to pick any specially defined signal.
Suppose we are mainly interested in regions with copy number equal to 1,
i.e., hemizygous deletion. Such deletion region should exhibit two
features: (1) log-ratio is equal to $\log(1/2) = -$0.693147 
(as versus $\log(2/2)=0$ in the normal situation); (2) homozygosity indicator 
equal to 1 (as versus to a mixture of $-$1's and 1's). 
For a SNP, we then define a ``one deletion" indicator variable whose
value is 1 if $-2 <$ log-ratio $< -0.34657$ (mid-point between $-0.693147$ and 0) 
and if its genotype is a homozygote, and the value $-1$ otherwise.

Figure 3 shows the cumulative plot for ``one deletion" indicator variable,
without or with detrending (by the linear trend in
the normal sample).  In both versions, the hemizygous deletion region 
can be seen clearly.  Not only the cumulative plot detects the CNA
region easily, but also it delineates the border of the deletion region 
accurately.

When deletion occurs in both chromosomes, called homozygous deletion,
the copy number is equal to zero. For homozygous deletions, both 
A- and B-channel intensity (see Methods) is close to zero, and the 
log($r$) is a large negative value. Because in the A- and B-channel
plane (see Methods), these SNPs are near the origin, the
angle $\theta$ can not be determined unambiguously. This leads to a
broad distribution of $\theta$ values between 0 and 1, as can be seen from Figure 1 (top).

We define a ``two deletions" indicator variable whose value is 1 
if the log-ratio is $< -2$; and the value is $-1$ otherwise. Note that 
the genotype information is not used. Figure 4 shows the cumulative 
plot for the ``two deletions" indicator variable for chromosome 13. 
One homozygous deletion region with $\sim$ 1Mb is clearly 
identified immediately adjacent to the 9Mb hemizygous deletion region.

The 9Mb deletion on chromosome 13 in our CLL sample, which was one 
of the known common deletions for this disease\cite{del13}, represents an
example of easy detection of CNA/CNV region, because the difference
between the normal and cancer cell for both log-ratio and genotype
sequence is already obvious from the raw data (Figure 1). The advantage
of cumulative plot is perhaps its ability to detect CNA/CNV region
of smaller sizes.

Figure 5 shows the example of chromosome 6 of our sample where 
there is no large-scaled CNA region. The log-ratio  and genotype
sequence look almost identical between the normal and the cancer 
cell. The cumulative plot for the ``one deletion" indicator 
variable shows that there are +400 more SNPs in the cancer 
cell than in the normal cell to have the one deletion signal
(the ``two deletion" cumulative plot is not shown because
the signal is mostly absent along the chromosome).
However, these SNPs are distributed throughout the chromosome, instead
of forming clusters,  and we still do not have strong evidence
that the cancer sample has more micro deletion regions as
compared to the normal sample.

In order to explore the possible existence of smaller CNA regions, 
we pick the longest ROH region (roughly 4Mb) and view it with 
cumulative plots. Figure 6 (left) shows the un-detrended
cumulative plot for the one-deletion indicator variable
in this region. A clear hemizygous deletion region should
show up as a jump in the cumulative plot. However,
the tendency within this ROH region is downward instead of upward. 
In other words, although all genotypes in this region are 
homozygous, the log-ratio mostly fails the $< -0.34657 $ criterion.

The failure in detecting hemizygous deletion at the Mb scale
does not necessarily prevent its possible existence at 
a smaller length scale. The right panel of Figure 6 shows a
200kb sub-region (marked in Figure 6 (left)) that contains
a 36kb region with an upward trend in the cumulative plot. 
A zooming into any small region in a cumulative plot enables 
it to detect CNA/CNV regions with ever smaller sizes.

It was previously suggested  that run of homozygosity can be a sequence
feature that is associated with certain human diseases\cite{lencz}.
We see here that ROH is only a partial indicator for a CNA/CNV region. 
The longest ROH on chromosome 6 in our sample only shows some
weak evidence in a much narrower region for one-deletion CNA. Considering 
both ROH and log-ratio sequence is clearly better than considering
ROH alone. Although ROH may still be biologically meaningful, as
it could reflect a copy-neutral loss-of-heterozygosity event, 
one has to obtain extra evidence to exclude population genetics events
such as inbreeding as the true cause. 

The pairing of the normal and the cancer sample is not essential
to our method. In Figures 3,4,6, the CNA regions can be identified
by cancer sample (the blue curve) alone. However, the comparison
with the normal sample provides supporting evidence that 
deletion only occurs in the cancer cell and not in the normal cell.
 
When SNPs along a chromosome are not evenly distributed, it may 
not be appropriate to move one step per SNP in the cumulative plot. 
For example, if multiple SNPs are in strong linkage disequilibrium in 
a densely typed region, the indicator variable values are positively 
correlated, and a sequence of +1 values is partially a consequence of 
their correlation, not as a series of independent evidences for CNA/CNV. 
We can adjust for this correlation by calculating the probability ratio $\alpha$ 
(see Method) in favor for concordant genotypes between neighboring 
SNPs, as compared to the average. If $\alpha > 1$, we discount 
a +1 (or $-1$) movement by dividing the $\alpha$ value.
For the chromosome 13 data, $\alpha$ is in a very narrow
range of (0.9921, 1.0002).
Because the probability ratio in favor of concordant homozygotes
is so close to 1, the adjusted cumulative plot is indistinguishable
from the original cumulative plot.

So far the delineation of an upward trend in the cumulative
plot is determined by visual inspection. Segmentation programs
can be developed to carry out the delineation automatically.
In particular, one may move along the cumulative plot, calculate
the slope from the start point to the moving position, then
from the moving position to the end point.  The position that 
maximizes the difference of the two slopes is chosen, leading
to the first segmentation. This segmentation can be carried out
recursively similar to the method described in \cite{pedro}.  

Finally, for case-control analysis using CNV, one deals with 
two groups of samples\cite{barnes}. In this situation, 
cumulative plot can be first applied to each individual person 
to identify the CNV/CNA region. Then, chromosomes can be 
partitioned into equal-sized windows and the frequency
of CNV/CNA-containing window in the case group is compared to
that in the control group for a statistical test.

\section*{Conclusions}

\indent

We have shown here that cumulative plots of an indicator 
variable derived from the log-ratio and SNP genotype sequence
can easily identify CNV or CNA regions. We illustrate the procedure
for hemizygous deletion (copy number equal to 1) and homozygous
deletion (copy number equal to 0) using samples taken from
a chronic lymphocytic leukemia patient. Although CNV/CNA regions at 
the Mb scale can also be detected by viewing the raw data, cumulative
plot is able to delineate the borders with higher degree of accuracy.
Another advantage of cumulative plot is perhaps in detecting
smaller CNV/CNA regions, such as those in the range of 10kb-100kb, 
as it is a scale-free approach that does not require a fixing of 
the window size. Cumulative plot is simple enough that no special-purpose
program is needed for its use except a graphic routine: for example, 
all results shown here are obtained by the general statistical 
package R\cite{r}.

\section*{Methods}

\subsection*{log-ratio and genotype data}

In a two-channel (two-color) SNP genotyping microarray, the A-
and B- channel (A- and B-allele) intensity reading is recorded.
These two intensities are normalized by reference intensity
values which are obtained by averaging many normal samples.
Each SNP can be represented by a point in the ($x,y$) plane
where $x,y$ are the normalized A- and B-channel intensity.
The polar coordinate of the point is $r=\sqrt{x^2+y^2}$
and $\theta= tan^{-1} (y/x)$\cite{snp-cgh}. Log(r) is the
``log ratio" value that provides a copy-number information, and
$\theta$ provides a genotype information, where $\theta=0,1$
correspond to two homozygotes, and $\theta = 0.5$ corresponds
to the heterozygote. Note: (1) $r$ value
depends on a group-averaged reference level, and this information
is provided by the array-maker company. (2) Although $r$
and $\theta$ is in principle independent, there could
be weak correlation between them.  Our starting point are
the two sequences of log(r) and discretized $\theta$ values
(i.e. genotype) along a chromosome. 

\subsection*{Cumulative plots for log-ratio and homozygosity sequence}

The $r$ and $\theta$ variable is transformed by: log-ratio
$= log(r)$ and homozygosity indicator $h= 4 \times |\theta-0.5| -1$.
For heterozygotes, $h$ is close to $-1$, and for two homozygotes, 
$h$ is close to 1. Denote the $i$-th SNP's log-ratio and homozygosity indicator
as $\log(r_i)$ and $h_i$. The (original) cumulative plots
of these two sequences are:
\begin{eqnarray}
 cumu.log.ratio_j &=& \sum_{i=1}^j \log(r_i) \nonumber \\
 cumu.h_j &=& \sum_{i=1}^j  h_i 
\end{eqnarray}

A cumulative plot can be detrended such that the first and the last
SNP are on the same horizontal line. The purpose of this detrending
is to remove the chromosome-wide bias so that regional deviations
are highlighted. In our normal and cancer cell from the same
individual example, we detrend the normal sample by subtracting 
the linear function $a+b x_i$, where $x_i$ is the Mb position of
the $i$th SNP, $N$ is the number of SNPs, and
\begin{eqnarray}
 b &=& \frac{cumu.log.ratio_N - cumu.log.ratio_1 }{ x_N- x_1} \nonumber \\
 a &=& cumu.log.ratio_N - b x_N.
\end{eqnarray}
To highlight the difference between the cancer cell and the normal
cell, we use the $a$ and $b$ obtained from the normal cell to
detrend the cumulative plot for the cancer cell.

\subsection*{Cumulative plots corrected by spacing between neighboring SNPs}

When SNPs are not distributed evenly along a chromosome, one may 
consider correcting the effect of inhomogeneous correlation between 
neighboring SNPs. We first calculate the probability of a neighboring
SNP of a homozygous SNP to be also homozygous due to the
correlation between them. This calculation is carried out
by the Haldane's map\cite{jurg}.

Haldane's map relates the number of recombinations within a chromosomal
interval $M$ and the probability of observing a recombinant between
the two end points $R$:
\begin{equation}
\label{eq4}
R = \frac{1-exp(-2M)}{2}.
\end{equation}
The unit of $M$ is Morgan, which is roughly equal to 100Mb
(or 1 centi Morgan is equal to 1 Mb\cite{ulgen}). The probability
of observing a non-recombinant is $1-R$.

Denote $p_{same}$ the probability that one homozygous SNP is followed by
another homozygous SNP that is $M$ genetic distance apart. Since
 Haldane formula is applicable to haplotype, or a single copy
of a chromosome, for two copies of a chromosome,
$p_{same}= (1-R)^2 \approx 1-2R = e^{-2M}$.

Suppose the average spacing between two neighboring SNPs is $\overline{M}$.
For a neighboring SNP pair whose spacing $M < \overline{M}$,
it is more likely for both SNPs to be homozygous than the average,
by a probability ratio of $\alpha = p_{same}/\overline{p}_{same}= e^{-2(M-\overline{M})}$,
and the cumulative plot for the homozygosity indicator variable
can be adjusted by dividing that ratio:
\begin{equation}
cumu.h_j = \sum_{i=1}^j h_i/\alpha_i = \sum_{i=1}^j h_i e^{2(M_{i-1,i}-\overline{M})}.
\end{equation}
We assume that $p_{same}$ is calculated in the same way as for other indicator
variables, meaning CNV/CNA of a particular type is maintained at the neighboring SNP
by the same probability $e^{-2M}$, and the above formula can be used
to correct other cumulative plots. Note that transition
probability from one genotype in a SNP to another genotype in the
neighboring SNP can also be estimated from the HapMap data.

\section*{Authors contributions}

W.L. designed the method, carried out the analysis, and wrote
the manuscript; A.L. genotyped the samples; P.K.G. proposed the 
CNV study of chronic lymphocytic leukemia.  

\section*{List of abbreviations}

{\bf CGH:} comparative genomic hybridization
{\bf CLL:} chronic lymphocytic leukemia
{\bf CNA:} copy number alterations
{\bf CNV:} copy number variations
{\bf GC\%:} guanine and cytosine contents
{\bf HMM:} hidden Markov models
{\bf ROH:} run of homozygosity
{\bf SNP:} single nucleotide polymorphism

\section*{Acknowledgements}

We thank Nick Chiorazzi for providing the CLL sample,
and Pedro Bernaola-Galv\'{a}n, Jos\'{e} Oliver for
discussions on segmentation methods.


\newpage


\section*{Figures}

  \subsection*{Figure 1 - Log-ratio and genotype sequences for chromosome 13
in paired samples from a CLL patient}
Log-ratio (top) and genotype $\theta$ (bottom) sequence for SNPs
from chromosome 13 of two samples taken from the same cancer patient:
black for normal cell and blue for cancer cell. For the log-ratio plot,
the copy number of 2 level $\log(2/2)=0$ and the copy number of 1 level
$\log(1/2)= -0.693147$  are marked.

\begin{figure}[t]
  \begin{turn}{-90}
   \epsfig{file=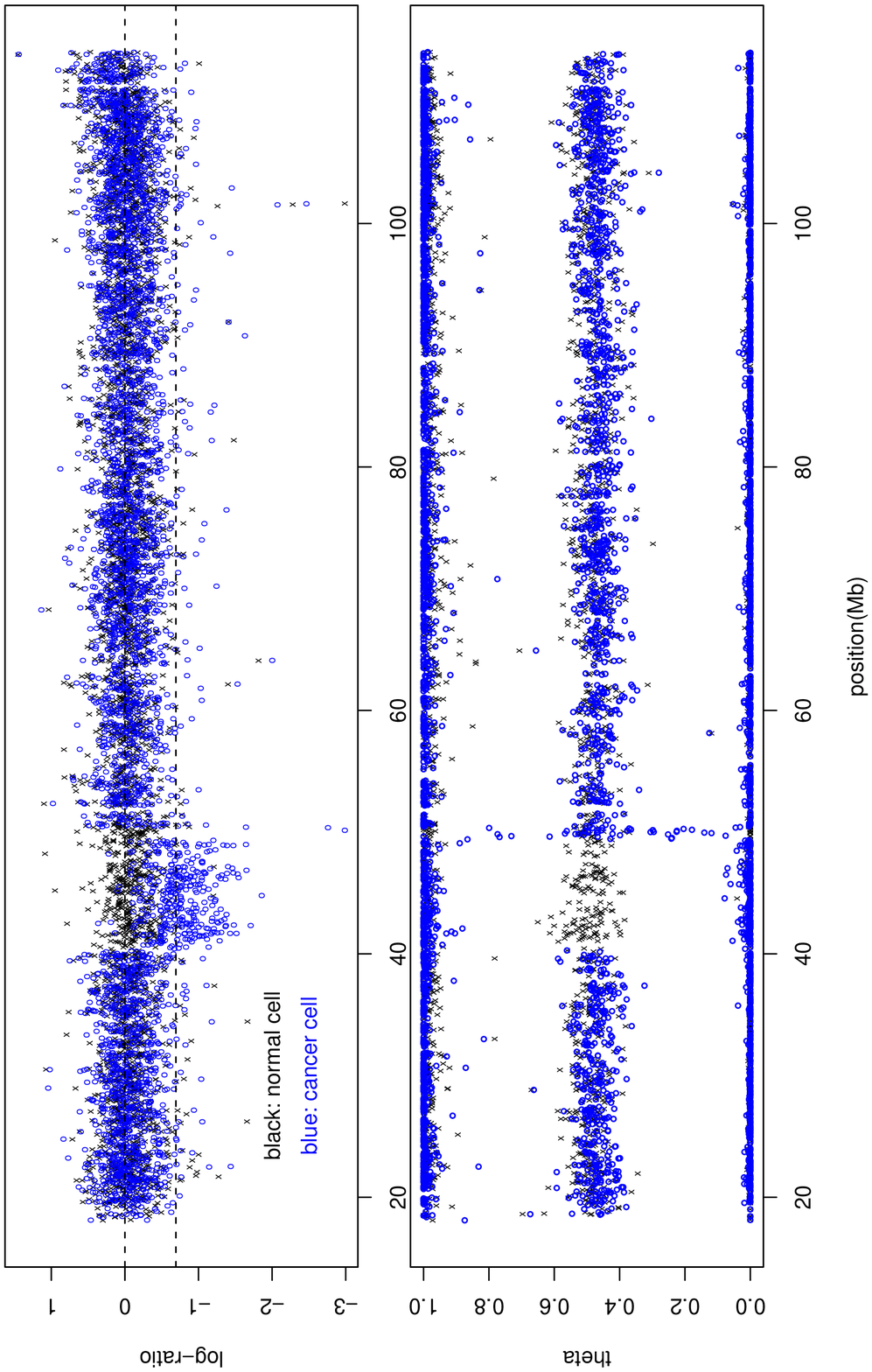, width=11cm}
  \end{turn}
\end{figure}

\newpage

  \subsection*{Figure 2 - Cumulative plot of log-ratio and homozygosity
sequence}
Cumulative plot and detrended cumulative plot for both the log-ratio
sequence and the homozygosity indicator sequence (for the chromosome 13
data shown in Figure 1).  Top: cumulative plots for log-ratio sequence.
Bottom: cumulative plot for homozygosity sequence (1 for homozygote,
$-1$ for heterozygote). Left: original cumulative plots. Right: detrended
cumulative plots. The linear trend obtained from the normal sample is
used to detrend both the normal and the cancer sample. Black for the normal
cell and blue for the cancer cell. The 9MB hemizygous deletion and the
neighboring 1Mb homozygous deletion region are marked by red lines.

\begin{figure}[t]
  \begin{turn}{-90}
   \epsfig{file=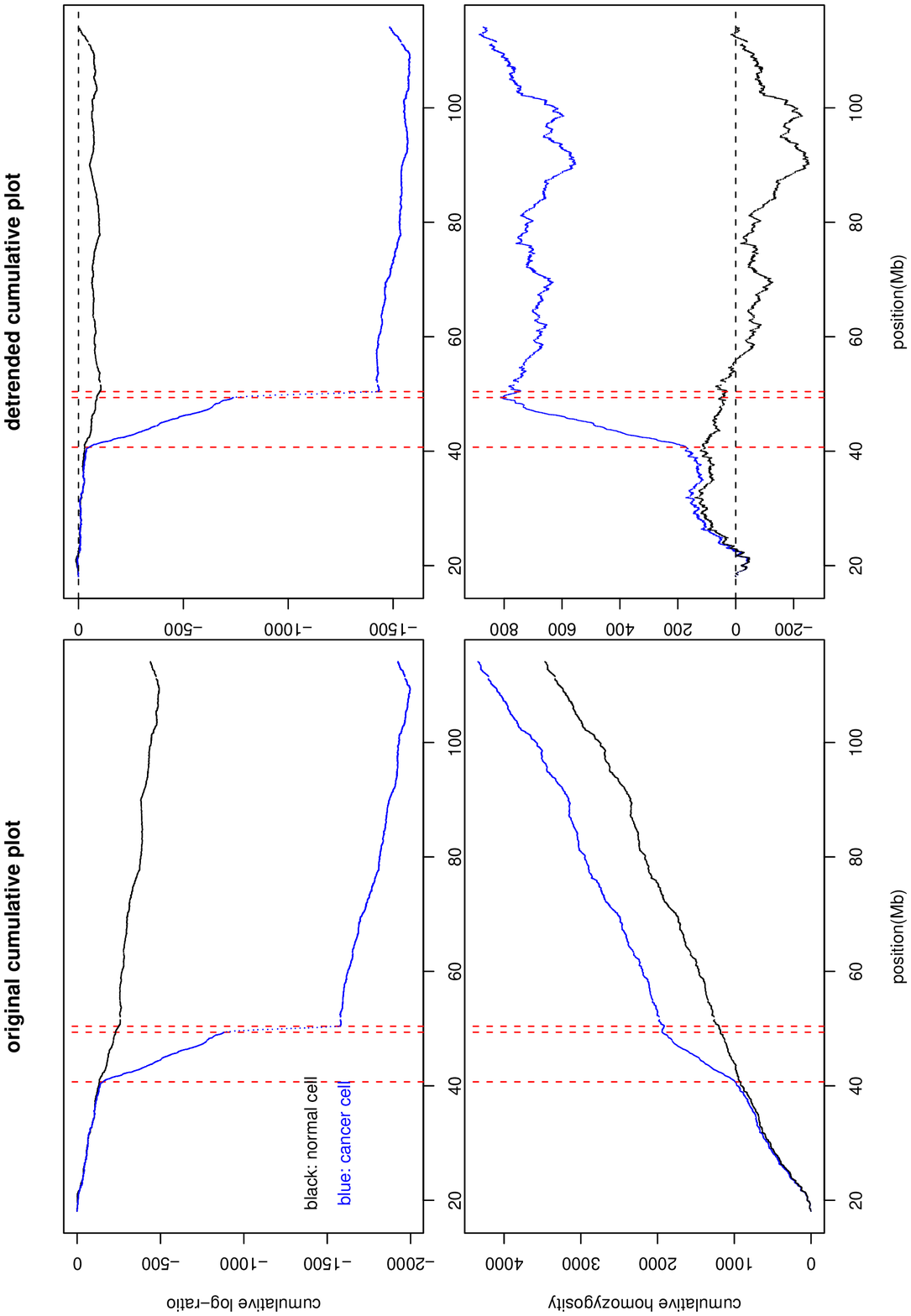, width=11cm}
  \end{turn}
\end{figure}

\newpage
  \subsection*{Figure 3 - Cumulative plot of the hemizygous deletion
indicator variable }
Cumulative plot (top) and detrended cumulative plot (bottom)
for the 9Mb hemizygous deletion region on chromosome 13, using
the ``one deletion" indicator variable.

\begin{figure}[t]
  \begin{turn}{-90}
   \epsfig{file=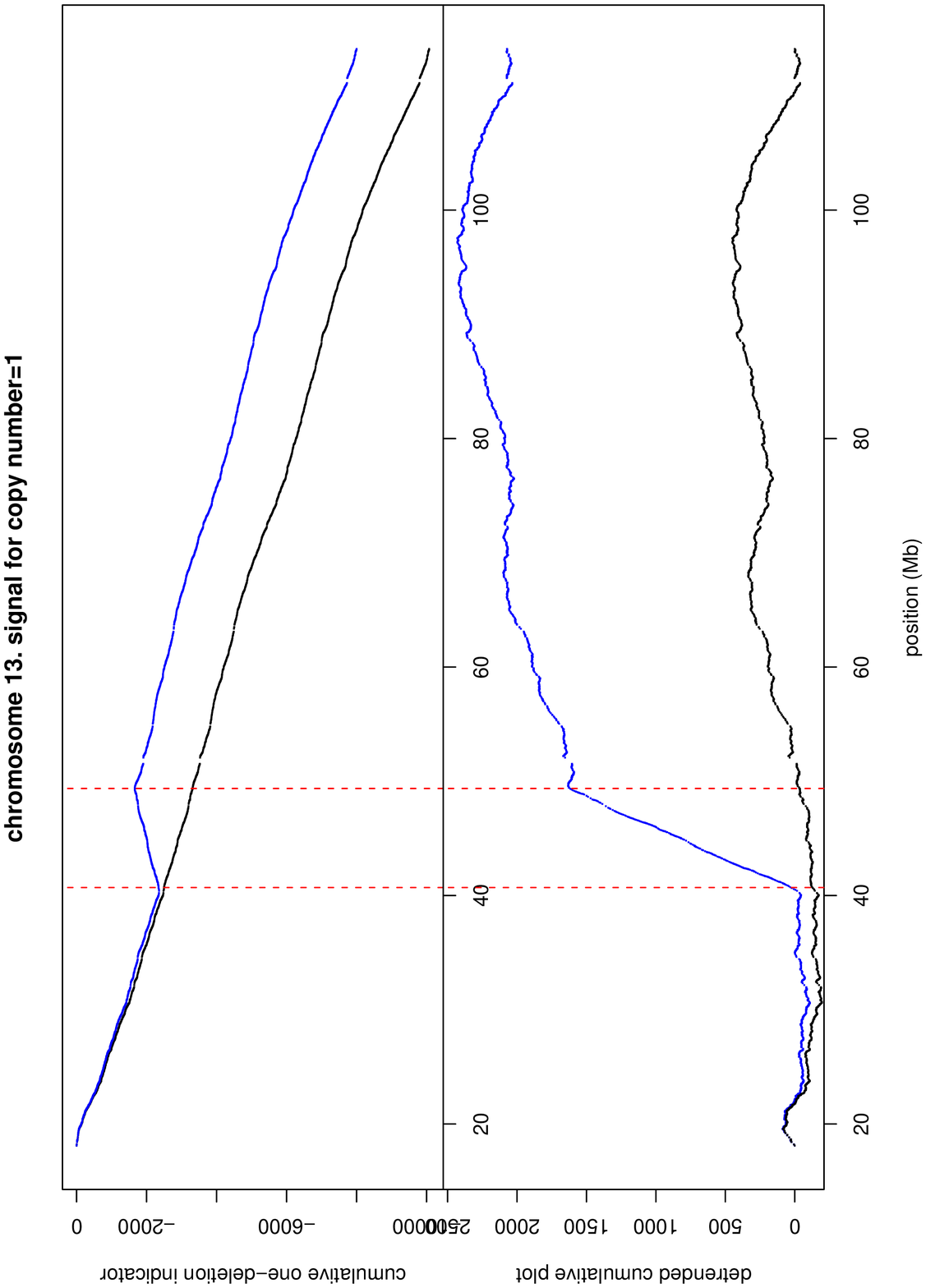, width=11cm}
  \end{turn}
\end{figure}

\newpage
  \subsection*{Figure 4 - Cumulative plot of the homozygous deletion
indicator variable }
Cumulative plot (top) and detrended cumulative plot (bottom)
for the 1Mb homozygous deletion region on chromosome 13, using
the ``two deletions" indicator variable.

\begin{figure}[t]
  \begin{turn}{-90}
   \epsfig{file=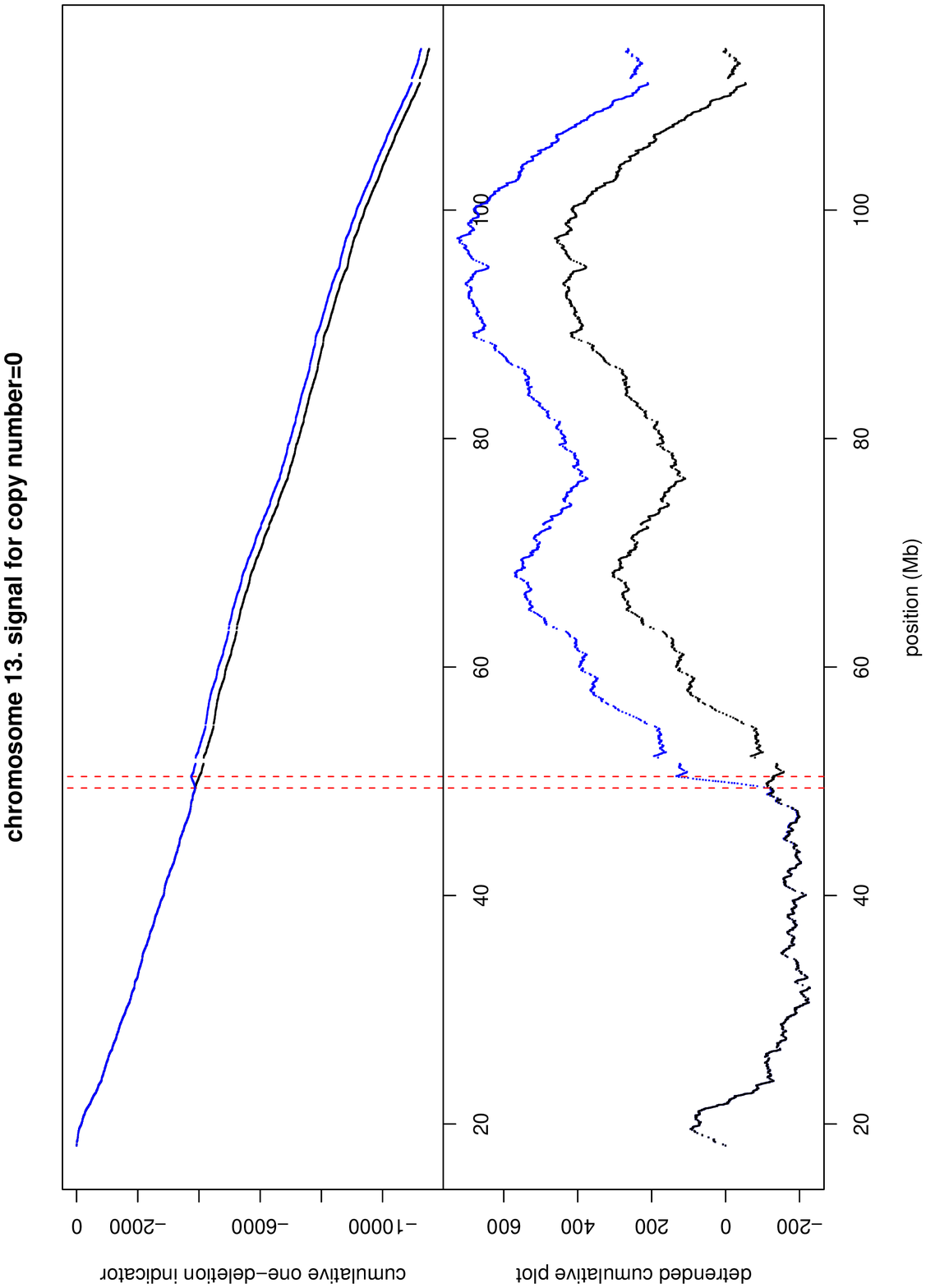, width=11cm}
  \end{turn}
\end{figure}

\newpage
  \subsection*{Figure 5 - Log-ratio and genotype sequences for
chromosome 6 in paired samples from a CLL patient}
The log-ratio sequence (top), genotype $\theta$ sequence (bottom),
and the detrended cumulative plot for the ``one deletion" indicator variable
for SNPs on chromosome 6. Black and blue color refer to the normal
and cancer cell sample taken from the same cancer patient. The largest
run-of-homozygosity region is marked by red vertical lines.
The copy number of 2 level $\log(2/2)=0$ and the copy number of 1 level
$\log(1/2)= -0.693147$  are marked in the log-ratio plot.

\begin{figure}[t]
  \begin{turn}{-90}
   \epsfig{file=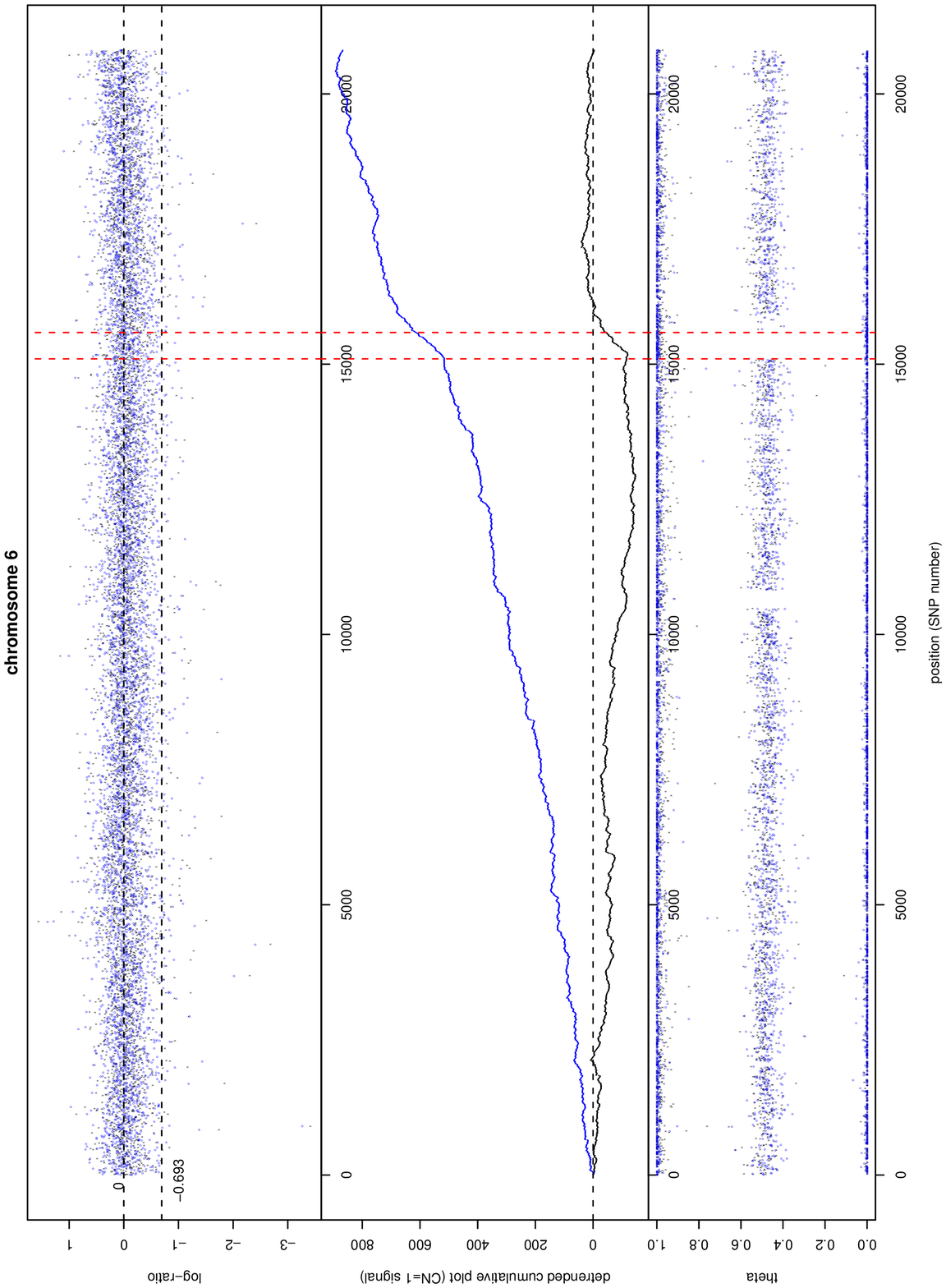, width=11cm}
  \end{turn}
\end{figure}
\newpage
  \subsection*{Figure 6 - Zoom in of smaller regions}
Cumulative plots of ``one deletion" indicator variable
for the region marked in Figure 5 (left), and the sub-region marked
by a horizontal bar on the left (right). Black and blue refer to
the normal and the cancer sample.

\begin{figure}[t]
  \begin{turn}{-90}
   \epsfig{file=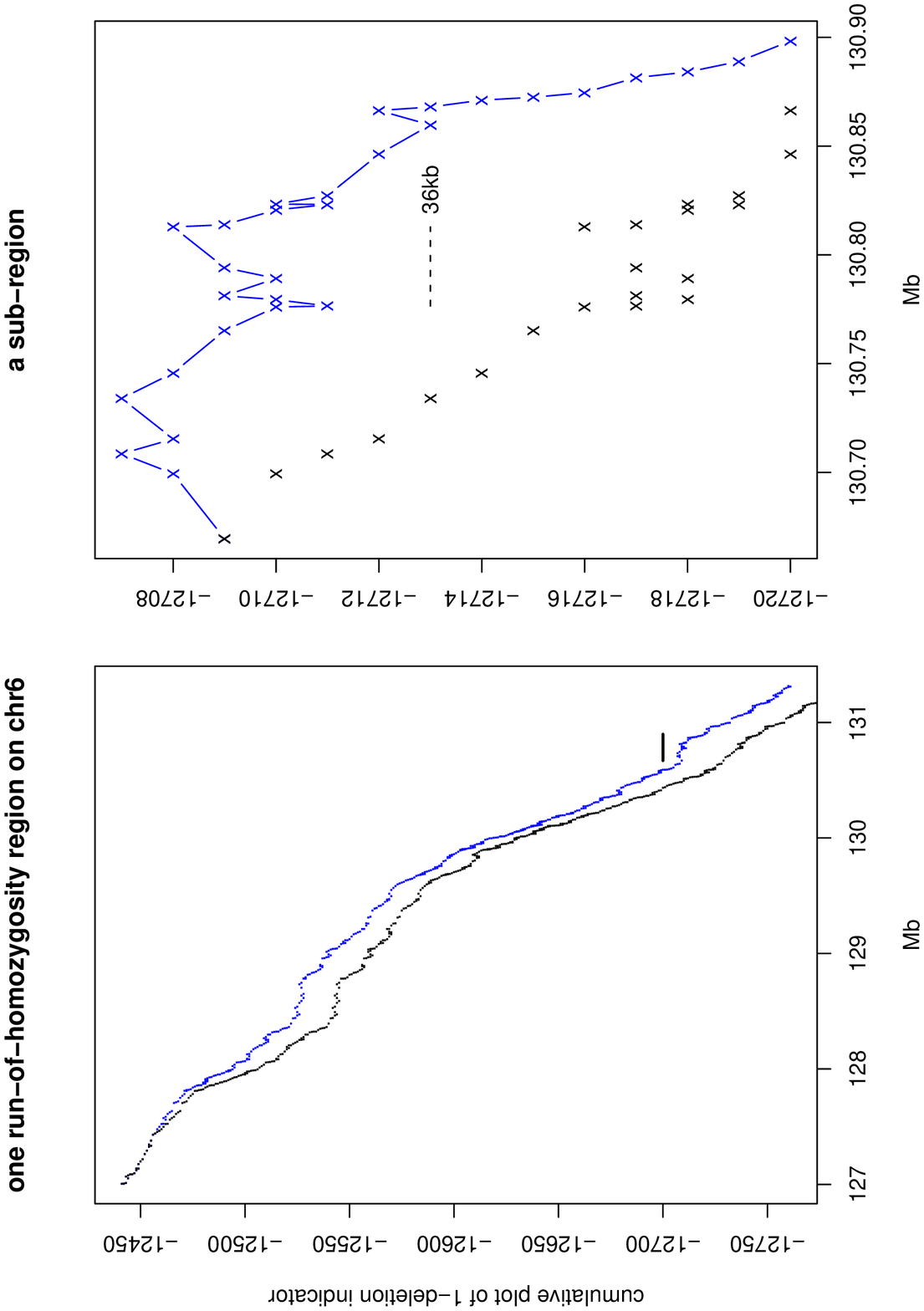, width=11cm}
  \end{turn}
\end{figure}

\end{document}